\pdfoutput=1
\documentclass[11pt,a4paper]{article}
\usepackage[T1]{fontenc}
\usepackage[utf8]{inputenc}
\usepackage{mathptmx}
\usepackage[scaled=0.92]{helvet}
\usepackage[margin=2.2cm]{geometry}
\usepackage{graphicx}
\usepackage{multirow}
\usepackage{amsmath,amssymb}
\usepackage{xcolor}
\usepackage{textcomp}
\usepackage{booktabs}
\usepackage{tabularx}
\usepackage{array}
\usepackage[font=small,labelfont=bf,labelsep=space]{caption}
\usepackage{microtype}
\usepackage{xurl}
\usepackage[hidelinks]{hyperref}

\newcommand{\repourl}{\url{https://doi.org/10.5281/zenodo.23620709}}

\newcommand{\tablebodyfont}{\footnotesize}
\newcommand{\TCH}[1]{\textbf{#1}}
\newcommand{\botrule}{\bottomrule}
\newcommand{\bmhead}[1]{\par\medskip\noindent\textbf{#1.}\ }
\newcolumntype{L}[1]{>{\raggedright\arraybackslash}p{#1}}
\newcolumntype{C}[1]{>{\centering\arraybackslash}p{#1}}

\title{\textbf{Trust Mechanisms for Offline Generative AI at the Edge: A Connectivity-Regime Framework and an Empirical Study on a Raspberry Pi~5}}
\author{%
  {\fontencoding{OT1}\fontfamily{cmr}\selectfont Georgios Feretzakis and Alexandros Papaspyridis}\\[1em]
  {\normalsize\fontencoding{OT1}\fontfamily{cmr}\selectfont securagen.ai}\\
  {\normalsize\fontencoding{OT1}\selectfont\href{mailto:research@securagen.ai}{\textcolor{magenta}{\texttt{research@securagen.ai}}}}%
}
\date{}

\begin{document}
\maketitle

\begin{abstract}
Generative AI and large language models (LLMs) are moving into the Internet of Things, yet most deployments assume cloud connectivity, and many IoT settings have intermittent, restricted, or no connectivity. How to make a local generative assistant trustworthy when it is disconnected, and what trust costs on edge hardware, remain largely unmeasured. We propose a connectivity-regime framework (C0--C4) that treats disconnection as a design dimension, a taxonomy of inference-time trust mechanisms, and signed, versioned knowledge packs as the update channel for disconnected regimes. We then evaluate these ideas on a Raspberry Pi~5 running fully local retrieval-augmented generation under network isolation, with three Qwen2.5 model sizes (1.5B, 3B, and 7B parameters), three context sizes, and 300 SQuAD~2.0 questions that are answerable, unanswerable, or outside the corpus. An empirically calibrated retrieval-score gate detects knowledge gaps almost perfectly (AUROC 0.97) but in-topic unanswerable questions poorly (AUROC 0.68). A deterministic output validator converts malformed outputs into explicit abstentions, lowering the risk among answered queries from 0.81 to 0.46 for the smallest model with no observed loss of correct answers; the NLI-based entailment verifier we tested adds little, and for the 7B model 96\% of substantive answers to unanswerable-labelled questions match normalized spans of the retrieved passages. The 3B model achieves the same gold-answer containment rate as the 7B model (75\%) at less than half the latency and estimated energy, although the 7B model far more often exceeds a token-F1 threshold of 0.5 (74\% versus 31\%) and has lower risk; five retrieved passages lower correctness from 75\% to 52\% compared with three. Verifying a signed knowledge pack takes about 4.5~ms, versus about 29~min for embedding its passages on the device; signature and hash checks alone accept six of nine tested update attacks that layered verification of version state, expiry, revocation, and compatibility rejects. Yet a validly signed poisoned pack passes every pack check: on 30 questions previously answered correctly, correct answers drop to zero and 14 return the injected answer; 13 of these pass gating, span, and entailment checks. None of the mechanisms we evaluated establishes the truth of the evidence. We derive design guidelines for trustworthy generative edge systems and release code and data.
\end{abstract}


\section{Introduction}\label{sec:intro}

Large language models (LLMs) are moving from cloud services into the Internet of Things (IoT) and the intelligent environments built on it~\cite{r1,r2}. Maintenance assistants, clinical document tools, and decision-support agents increasingly mediate access to operational knowledge. Yet the dominant deployment pattern remains cloud-centric. That pattern fails exactly where many IoT deployments operate: maritime, remote-infrastructure, disaster-response, industrial, and defense settings with intermittent, restricted, or deliberately absent connectivity~\cite{r3}. Data-protection and AI regulation~\cite{r4,r5,r6} add further pressure to keep inference within the operator's perimeter. And the propensity of LLMs to produce fluent but unsupported content~\cite{r7} is most dangerous offline, where users cannot cross-check an answer.

Running a generative assistant locally is now feasible. Small models, 4-bit quantization, and portable runtimes bring billion-parameter models to embedded boards~\cite{r8,r9,r10,r11}, and retrieval-augmented generation (RAG) lets a local model answer from a local corpus~\cite{r12}. Prior work has benchmarked small language models for edge deployment~\cite{r13} and LLM inference on single-board computers~\cite{r14}, co-designed edge models for latency and energy~\cite{r15}, fitted RAG into edge memory budgets~\cite{r16}, and demonstrated offline RAG over Bluetooth mesh networks~\cite{r17} and offline-first edge architectures~\cite{r18}. Local and offline RAG is therefore not new in itself. What is far less understood is how to make such a system \emph{trustworthy} when it is disconnected, and what trust costs on edge hardware. Recent surveys of edge and on-device LLMs focus on efficiency~\cite{r19,r20,r21,r22,r23,r24}, and treat connectivity as background rather than as a design dimension. Trust mechanisms for RAG, such as selective answering, conformal risk control, and unanswerability evaluation~\cite{r25,r26,r27,r28,r29,r30}, have been proposed and evaluated on server-class hardware with connected evaluation pipelines. Their behavior, their interplay, and their cost on a CPU-only edge device under air-gapped conditions remain largely unmeasured.

This paper addresses that gap with a framework and an empirical study. We ask five research questions:
\begin{itemize}
\item \textbf{RQ1.} How well does a generator-independent retrieval-score gate separate queries that should be answered from knowledge-gap queries and from in-topic unanswerable queries?
\item \textbf{RQ2.} How do layered trust mechanisms (generator self-abstention, deterministic output validation, calibrated score gating, and post-hoc entailment verification) interact, and what does each cost in correct answers?
\item \textbf{RQ3.} How do generator size and the number of retrieved passages trade answer quality against latency and energy on a CPU-only edge device?
\item \textbf{RQ4.} What does verifying and activating a signed knowledge pack cost on the device, and which update attacks do signature-and-hash checks and layered verification stop?
\item \textbf{RQ5.} Can a validly signed but poisoned knowledge pack subvert the system, and do score gating or post-hoc verification catch it?
\end{itemize}

Our contributions are as follows:
\begin{enumerate}
\item A \textbf{connectivity-regime framework} (C0--C4) that treats disconnection as a spectrum from cloud-connected to air-gapped operation, and traces its consequences for inference, updates, federation, threats, and governance evidence (Sect.~\ref{sec:framework}).
\item A \textbf{taxonomy of inference-time trust mechanisms}, organized by locus and dependence on the generator, and a comparison of \textbf{federation units} that motivates signed, versioned \emph{knowledge packs} as the update channel for disconnected regimes (Sect.~\ref{sec:framework}).
\item An \textbf{empirical evaluation} of these mechanisms on a Raspberry Pi~5 under network isolation: five configurations, three model sizes, three context sizes, and 1,500 generations over 300 questions, plus 30 generations on a poisoned pack (Sects.~\ref{sec:method} and~\ref{sec:results}). Score gating detects knowledge gaps almost perfectly (AUROC 0.97) but in-topic unanswerable questions poorly (0.68). A deterministic output validator converts malformed outputs into explicit abstentions, with negligible overhead and no observed loss of correct answers. A 3B model achieves the same gold-answer containment rate as a 7B model at less than half the latency and energy, although the 7B model far more often exceeds a token-F1 threshold of 0.5 and has lower risk. More retrieved context can lower answer quality. Signature-and-hash checks alone accept six of nine tested update attacks that a layered verifier rejects. Yet a validly signed poisoned pack passes every pack check and eliminates correct answers on the targeted questions, and most injected answers also pass score gating and the post-hoc checks we tested.
\item \textbf{Design guidelines} for trustworthy generative edge systems and an open release of code and data (Sect.~\ref{sec:discussion}).
\end{enumerate}

Section~\ref{sec:bg} reviews background and related work. Section~\ref{sec:framework} presents the framework. Sections~\ref{sec:method} and~\ref{sec:results} describe the experimental methodology and results. Section~\ref{sec:discussion} discusses design guidelines, governance evidence, threats to validity, and open problems, and Sect.~\ref{sec:conclusion} concludes.

\section{Background and Related Work}\label{sec:bg}

\subsection{Local LLM Inference at the Edge}\label{sec:bg-inference}

Edge intelligence places training and inference along the cloud--edge continuum~\cite{r31}, and TinyML extends it to microcontrollers~\cite{r32}. Generative workloads changed the profile: autoregressive models are dominated by billions of parameters and a growing key--value cache. Three developments made local generation feasible. Capable small models were designed for on-device use~\cite{r8,r33}. Compression reduced their footprint, through post-training quantization~\cite{r10,r11}, pruning~\cite{r34}, distillation~\cite{r35}, and parameter-efficient adaptation~\cite{r36,r37}. Efficient serving runtimes span datacenter engines~\cite{r38} and portable CPU/GPU frameworks~\cite{r9}. Client-side benchmarks now measure local LLM performance across heterogeneous hardware~\cite{r39}. A study of more than 60 small language models found that model architecture, context processing, and compression jointly shape on-device cost and capability~\cite{r13}, and benchmarks on Raspberry Pi and Orange Pi boards show strong dependence on runtime, device, and model size~\cite{r14}. Algorithm--hardware co-design can yield large latency and energy savings~\cite{r15}. On the retrieval side, EdgeRAG prunes and regenerates embeddings to fit RAG into edge memory~\cite{r16}, offline RAG has been demonstrated over Bluetooth mesh networks~\cite{r17}, and practitioner architectures now describe offline-first generative AI at the edge~\cite{r18}. These works measure speed, memory, and retrieval quality; they do not measure grounding, abstention, or the cost of trust mechanisms.

\subsection{Retrieval Augmentation, Hallucination, and Abstention}\label{sec:bg-rag}

RAG couples a parametric generator with a non-parametric memory~\cite{r12}. Dense passage retrieval provides an efficient mechanism for selecting relevant evidence~\cite{r40}, and RAG has been surveyed extensively~\cite{r41}. Retrieval spans sparse ranking, dense bi-encoders, and multilingual embedding models~\cite{r42}, served by approximate nearest-neighbor indices~\cite{r43}. For disconnected deployment, RAG has a crucial property: it decouples knowledge from weights, so updating a corpus and its index is often the only practical way to keep an assistant current. RAG evaluation has been partly standardized through reference-free metrics~\cite{r44}, automated judges~\cite{r45}, and benchmarks for robustness and negative rejection~\cite{r46}, including Chinese~\cite{r47} and multilingual settings~\cite{r48}.

Hallucination is the central reliability challenge of generative models~\cite{r7,r49}. Retrieval reduces it but does not eliminate it~\cite{r50}. Complementary mechanisms include sampling-based self-consistency~\cite{r51}, self-knowledge probing~\cite{r52}, and reflective retrieve--generate--critique training~\cite{r53}. A structurally different family decides whether to answer at all: selective prediction~\cite{r54}, selective question answering under domain shift~\cite{r25}, and conformal methods with distribution-free guarantees~\cite{r55,r56}, surveyed for LLMs in~\cite{r57}. Recent work applies these ideas to RAG, through conformal calibration of retrieval scores~\cite{r26}, certified generation risk~\cite{r27}, and counterfactual risk control~\cite{r28}, and learns instance-adaptive conformal abstention policies instead of a single global threshold~\cite{r30}. UAEval4RAG makes unanswerability a first-class RAG evaluation target, with six categories of knowledge-base-specific unanswerable requests. It finds that no single RAG configuration performs best across datasets, and that rejection often stems from retrieval failure rather than from recognizing that a request should not be answered~\cite{r29}. Its \emph{Out-of-Database} category, covering requests that are highly relevant to the knowledge base but have no answer in it, corresponds most closely to the in-topic unanswerable questions in our study. Citation generation~\cite{r58} and factuality scoring~\cite{r59} support verification. We found no study that evaluates these mechanisms jointly, on edge hardware, under disconnection.

\subsection{Federation and Knowledge Distribution}\label{sec:bg-federation}

Federated learning (FL) trains shared models without centralizing raw data~\cite{r60}. It addresses statistical and systems heterogeneity~\cite{r61} and has been surveyed for IoT~\cite{r62}. For LLMs, federation centers on parameter-efficient adapters~\cite{r63,r64}. FL has well-known weaknesses: gradients leak data~\cite{r65}, protocols admit backdoor and Byzantine adversaries~\cite{r66}, and malicious aggregators can implant backdoors in federated LLM question answering~\cite{r67}. Federated RAG distributes the knowledge plane instead~\cite{r68}, through anonymized memories~\cite{r69}, federated vector databases~\cite{r70}, or parametric adapters~\cite{r71}. All of these presuppose recurring connectivity.

\subsection{Security and Governance of Local Generative AI}\label{sec:bg-security}

Local deployment removes the external inference provider from the data flow but creates its own attack surface~\cite{r72,r73}. Indirect prompt injection compromises LLM-integrated applications through processed content~\cite{r74,r75}, adaptive attacks bypass published defenses~\cite{r76}, and community taxonomies consolidate these risks~\cite{r77}. Corpus poisoning steers RAG answers with a few crafted passages~\cite{r78,r79}, and certifiable defenses exist for bounded corruption~\cite{r80}. Upstream, web-scale training data can be poisoned~\cite{r81}, and malicious pickle-serialized model files evade hub scanners~\cite{r82}. Model signing~\cite{r83} and non-executable weight formats~\cite{r84} address provenance and code execution, respectively. Secure update frameworks go further. The Update Framework (TUF) separates signing roles and adds snapshot and timestamp metadata against rollback and freeze attacks~\cite{r85}. The IETF manifest information model for IoT firmware updates requires a monotonic sequence number against rollback and allows an expiration time~\cite{r86}. In air-gapped settings, covert channels~\cite{r87} and removable media dominate the threat model. On the privacy side, membership inference~\cite{r88} and training-data extraction~\cite{r89} make both weights and retrieval corpora privacy-relevant; differential privacy~\cite{r90}, trusted execution~\cite{r91}, and secure aggregation~\cite{r92} are the classical defenses.

Governance instruments relevant to local generative AI include the GDPR~\cite{r4}; the EU AI Act~\cite{r5} as amended by the Digital Omnibus on AI~\cite{r6}; the Cyber Resilience Act~\cite{r93}; the NIST AI Risk Management Framework and its Generative AI Profile~\cite{r94,r95}; and ISO/IEC 42001~\cite{r96}. Several AI Act obligations (record-keeping, human oversight, accuracy, robustness, and cybersecurity) attach only to high-risk systems. Following the Digital Omnibus, the requirements in Sections 1--3 of Chapter III, with the exception of Article 6(5), apply from 2 December 2027 for systems classified as high-risk pursuant to Article 6(2) and Annex III, and from 2 August 2028 for those classified pursuant to Article 6(1) and Annex I~\cite{r6}.

\subsection{Positioning}\label{sec:bg-positioning}

Recent surveys consolidate edge and on-device LLMs~\cite{r19,r20,r21,r22,r23,r24}, security-oriented LLM edge intelligence~\cite{r72,r97}, multi-LLM edge systems with trust~\cite{r98}, RAG security~\cite{r73}, federated RAG~\cite{r68}, LLMs for IoT ecosystems~\cite{r2}, and offline edge deployment~\cite{r99}. Our earlier work examined privacy-preserving techniques for generative AI and the protection of sensitive data in large language models~\cite{r100,r101}. Edge RAG, offline connectivity, selective answering, and RAG unanswerability have thus each been studied~\cite{r16,r17,r25,r29}. To our knowledge, however, no prior work couples a connectivity-regime framework with an empirical measurement of trust mechanisms (abstention, score gating, and output validation), of their latency and energy cost on edge hardware, and of signed knowledge activation under network isolation. This paper contributes that combination.

\section{A Connectivity-Regime Framework}\label{sec:framework}

\subsection{Connectivity Regimes C0--C4}\label{sec:regimes}

We treat connectivity as a first-class architectural dimension, because it determines which inference, update, federation, and assurance mechanisms are admissible. Table~\ref{tab5} defines five regimes, from cloud-connected (C0) to air-gapped (C4). The taxonomy is our proposal, not an established standard. Its value lies in tying each regime to the update channels, threats, and assurance mechanisms it admits. The regimes are deployment regimes rather than a strict connectivity scale: C0 and C1 may have the same connectivity and differ in where inference runs. We use \emph{air-gapped} in the strict sense of an interface with no physical connection, across which any transfer is manual and under human control~\cite{r102}; permanently installed unidirectional gateways (data diodes) are therefore placed in C3 rather than C4.

\begin{figure*}[!t]
\centering
\includegraphics[width=0.9\textwidth]{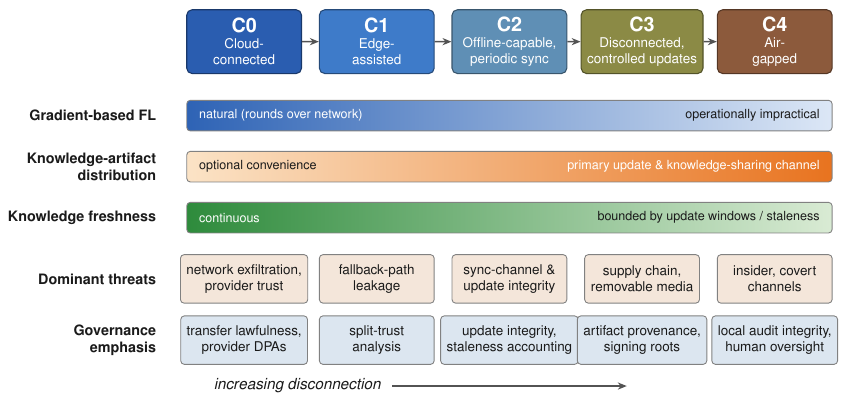}
\caption{The C0--C4 connectivity continuum and its architectural consequences: as deployments move from cloud-connected (C0) to air-gapped (C4), gradient-based federation---i.e., federated learning (FL)---loses viability, knowledge-artifact distribution becomes the primary update and knowledge-sharing mechanism (supporting federation where multi-node curation is present, per Sect.~\ref{sec:packs}), the dominant threat surface shifts from network exfiltration to supply-chain, insider, and covert-channel risks, and governance emphasis moves from transfer lawfulness to provenance and local auditability}\label{fig1}
\end{figure*}

Increasing disconnection constrains the feasible mechanisms and reorders threats (Fig.~\ref{fig1}). Gradient-based FL is natural at C1--C2, marginal at C3, and operationally impractical at C4. External knowledge becomes staler as updates shrink to controlled windows, which makes versioned corpora and explicit staleness accounting progressively more important. The attack surface inverts: network exfiltration dominates at C0--C1, whereas at C3--C4 the decisive risks are the update supply chain, removable media, insiders, and covert channels~\cite{r87}. Governance emphasis shifts accordingly, from the lawfulness of data transfers (C0--C1) to artifact provenance, local audit integrity, and human oversight (C3--C4).

\begin{table*}[!t]
\tablebodyfont\setlength{\tabcolsep}{4pt}\sloppy
\caption{The C0--C4 connectivity-regime taxonomy for edge AI deployments and its architectural consequences}\label{tab5}
\begin{tabularx}{\textwidth}{L{2.7cm} L{3.4cm} L{2.4cm} L{3.7cm} X}
\toprule
\TCH{Regime} & \TCH{Definition} & \TCH{Inference} & \TCH{Learning / updates} & \TCH{Trust emphasis} \\
\midrule
C0: Cloud-connected & Persistent broadband to cloud & Cloud or hybrid & Continuous; provider-managed & Transit privacy; provider trust \\
C1: Edge-assisted & Cloud reachable; latency/privacy push compute to edge & Local with cloud fallback & Frequent synchronization; FL feasible & Split trust; fallback-path leakage \\
C2: Offline-capable & Operates offline; periodic synchronization & Fully local & Scheduled sync windows; FL rounds possible & Update integrity; staleness bounds \\
C3: Disconnected, controlled updates & No routine bidirectional connectivity; updates via controlled channels/media & Fully local & Signed artifacts over controlled channels, including approved one-way (unidirectional gateway) links & Supply-chain integrity; provenance \\
C4: Air-gapped & No physical connection to external networks; transfers only through controlled manual procedures & Fully local; all assets pre-provisioned & Controlled offline transfer or removable media; verified before activation & Covert channels \cite{r87}; insider threat; auditability \\
\botrule
\end{tabularx}
\end{table*}

\subsection{Inference-Time Trust Mechanisms}\label{sec:trust}

Table~\ref{tab3} organizes inference-time trust mechanisms by locus, dependence on the generator, and type of assurance. They address different failure modes and are complements, not substitutes. Mechanisms that rely on the generator's self-assessment inherit its unreliability. Mechanisms grounded in generator-independent signals, such as retrieval-side scores, deterministic checks, external verifiers, and human judgment, provide a different type of assurance.

Two mechanisms are central to our evaluation. A \emph{retrieval-score gate} withholds generation when the top retrieved passage is insufficiently similar to the query. Its signal depends on the embedding model rather than on the generator, and it can be calibrated with conformal methods~\cite{r26,r55}. Its protection is conditional: an unsupported query can still be similar to superficially related passages. An \emph{output validator} applies deterministic checks to the generator's response, for example rejecting a response that carries citation markers but no answer text. It is the cheapest generator-independent mechanism, and we include it because small models frequently produce such degenerate outputs (Sect.~\ref{sec:res-layers}).

\begin{table*}[!t]
\tablebodyfont\setlength{\tabcolsep}{4pt}\sloppy
\caption{Inference-time trust mechanisms for generative language systems, organized by mechanism locus, dependence on the generator, and type of assurance provided}\label{tab3}
\begin{tabularx}{\textwidth}{L{3.6cm} L{3.4cm} X L{3.0cm}}
\toprule
\TCH{Mechanism} & \TCH{Locus} & \TCH{Assurance character} & \TCH{Representative work} \\
\midrule
Prompt-level constraints & Generator input & Behavioral tendency only & prompting practice; \cite{r41} \\
Retrieval grounding & Evidence supply & Reduces, does not eliminate, hallucination & \cite{r12,r50} \\
Citation grounding & Output format & Traceability of claims to sources & \cite{r58} \\
Output validation (deterministic) & Generator output & Generator-independent, rule-based; catches degenerate or malformed outputs, not unsupported content & This paper; Sect.~\ref{sec:res-layers} \\
Self-verification / self-critique & Generator itself & Model-dependent; inherits model unreliability & \cite{r51,r52,r53} \\
Retrieval-score gating & Retrieval-side statistic & Generator-independent routing signal (embedding-model-dependent; calibratable) & \cite{r25,r26}; Sect.~\ref{sec:res-gating} \\
Selective generation / abstention & Decision layer & Structural: generation withheld for queries routed to abstention & \cite{r25,r54,r55} \\
External factuality verification (post hoc) & Independent verifier against a knowledge source & Bounded by verifier and knowledge-source quality & \cite{r59} \\
Human-in-the-loop review & Human oversight & Expert review for high-stakes decisions; effectiveness depends on reviewer expertise and workflow; costly & \cite{r5,r94} \\
\botrule
\end{tabularx}
\end{table*}

\subsection{Knowledge Packs as the Update Channel for Disconnected Regimes}\label{sec:packs}

How should disconnected nodes receive new knowledge? FL answers with parameters: gradients, weight deltas, or adapters. RAG suggests a second answer: \emph{knowledge packs}, that is, curated documents with their vector index, distributed as versioned, integrity-protected bundles over whatever channel the regime permits, including controlled offline transfer. Table~\ref{tab6} compares the two planes. Parameter units resist human review, can leak training data~\cite{r65}, and have diffuse effects on behavior. Knowledge packs can be verified cryptographically at activation and revoked one by one, and because RAG grounds answers in their content, the evidence behind an answer is attributable to identifiable, versioned artifacts. Their limitation is that they distribute knowledge, not skills.

Securing a knowledge pack involves three distinct layers, and a signature covers only the first.
\emph{Authenticity and integrity}: a signature over a manifest of hashes shows that the bytes come from an authorized publisher and have not changed since signing~\cite{r83}.
\emph{Freshness and anti-rollback}: a valid but outdated pack must still be refused, which requires monotonic sequence numbers, expiry, and compatibility metadata, as in TUF and the IETF IoT update manifest~\cite{r85,r86}.
\emph{Content trust}: a correctly signed pack can still contain poisoned passages or injected instructions~\cite{r74,r78}, so curation, adversarial testing, and post-generation verification remain necessary.
A RAG-specific pack manifest should therefore also record the embedding model, the chunking and index-build configuration, source provenance, and the calibration set used for any score threshold.

This view implies an operational model for C3--C4 regimes. A publisher builds, curates, and signs packs off-device. Edge nodes verify signature, hashes, and version state before activation, and refuse anything that fails. The model assumes that verification is cheap and building is expensive on the device. RQ4 tests this assumption and compares signature-and-hash checks with layered verification across update attacks; RQ5 tests the content-trust layer with a validly signed poisoned pack.

\begin{table*}[!t]
\tablebodyfont\setlength{\tabcolsep}{4pt}\sloppy
\caption{Federation units for collaborative edge intelligence: comparative properties. Entries are the authors' qualitative synthesis; cited works substantiate the federation or retrieval mechanism itself rather than every listed property}\label{tab6}
\begin{tabularx}{\textwidth}{L{2.3cm} L{2.2cm} L{2.2cm} L{2.4cm} L{1.7cm} L{1.7cm} X}
\toprule
\TCH{Federation unit} & \TCH{Raw data leaves node?} & \TCH{Human-auditable?} & \TCH{Connectivity need} & \TCH{Update size} & \TCH{Revocable?} & \TCH{Air-gap (C4) suitable?} \\
\midrule
Full model weights & No (but extraction risk \cite{r89}) & No & High bandwidth & GBs & Version rollback only & Marginal (offline transfer) \\
Gradients / deltas (FL) \cite{r60} & No (leakage risk \cite{r65}) & No & Recurring rounds & MBs--GBs & Effectively no & Opera\-tionally impractical \\
PEFT adapters (LoRA) \cite{r36,r37} & No & No & Periodic for iterative exchange; none for one-way artifact transfer & MBs & Per-adapter & Possible via controlled transfer; iterative exchange impractical \\
Knowledge packs (docs + indices) & Not necessarily; depends on pack contents \& curation & Partially (docs/metadata yes; embeddings no) & Any, incl.\ controlled offline transfer & Content-propor\-tional & Per-pack & Yes \\
Knowledge-graph updates \cite{r103} & Not necessarily; depends on graph contents \& curation & Partially & Any & Small--medium & Per-update & Yes \\
Prompt/policy bundles & Not necessarily; depends on bundle contents \& curation & Yes & Any & KBs & Per-bundle & Yes \\
\botrule
\end{tabularx}
\end{table*}

\subsection{Reference Architecture and Governance Evidence}\label{sec:architecture}

\begin{figure*}[!t]
\centering
\includegraphics[width=0.92\textwidth]{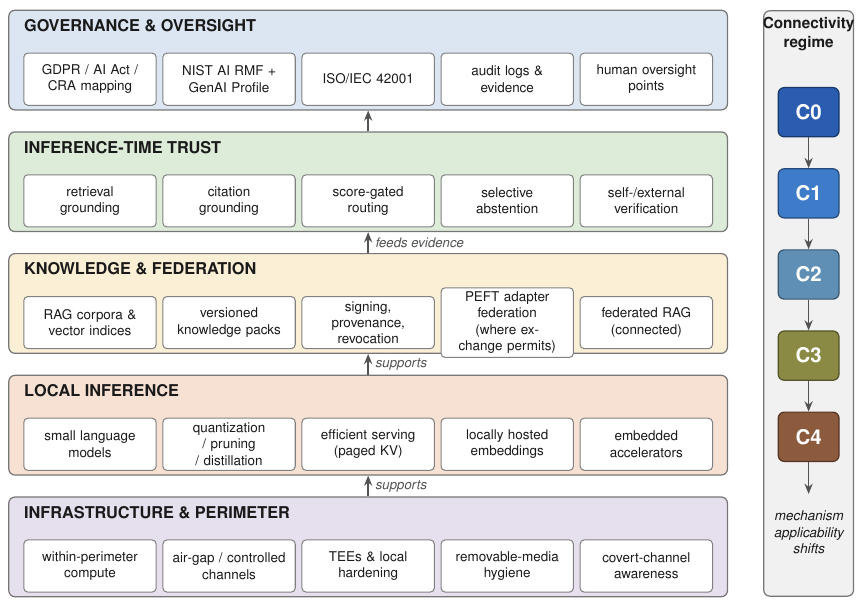}
\caption{A cross-layer map of trustworthy generative edge intelligence: infrastructure and inference mechanisms (bottom) support the knowledge and federation layer, whose artifacts feed inference-time trust mechanisms, all of which generate the evidence consumed by the governance layer (top). Mechanism applicability varies with the C0--C4 connectivity regime (right). GDPR: General Data Protection Regulation; CRA: Cyber Resilience Act; NIST AI RMF: NIST Artificial Intelligence Risk Management Framework; GenAI: generative artificial intelligence}\label{fig2}
\end{figure*}

Figure~\ref{fig2} composes these elements into a cross-layer reference architecture for C3--C4 regimes. A locally served, quantized generator~\cite{r9,r10} is paired with local multilingual embeddings and a graph-based index~\cite{r42,r43}. Answers are grounded in retrieved passages with citations~\cite{r12,r58} and routed by layered trust mechanisms to a full answer, a caveated answer, or an evidence-bearing abstention. Knowledge enters only through verified packs, and confidence signals, sources, and routing decisions are logged locally.

This logging is also what makes several governance obligations tractable to evidence without provider telemetry. Where they apply, AI Act record-keeping and human oversight for high-risk systems~\cite{r5,r6}, NIST AI RMF measurement and management functions~\cite{r94,r95}, and Cyber Resilience Act integrity requirements~\cite{r93} map onto pack versions, per-answer retrieval scores and validator outcomes, and routing logs. Observability does not by itself establish legal compliance, and parametric knowledge in the weights is not captured by such records.

\section{Experimental Methodology}\label{sec:method}

We instantiated a minimal version of the reference architecture and evaluated it under network isolation approximating the connectivity constraints of C4. The study is an illustration on one device and one dataset, not a benchmark.

\emph{Hardware and isolation.} All measurements used a Raspberry Pi~5 with a Broadcom BCM2712 (four Cortex-A76 cores), 16~GB RAM, 64-bit Raspberry Pi OS, and active cooling. Retrieval, the model server, and the benchmark ran inside a Linux network namespace that contained only the loopback interface, so the pipeline had no network access; an outbound request issued from inside the namespace failed, as expected. This demonstrates that the software pipeline has no network dependency. It does not reproduce the physical-security, removable-media, insider, or covert-channel threat model of an operational air gap. The full run took about 11.5~h, and no configuration recorded thermal throttling.

\emph{Pipeline.} The generators were Qwen2.5-Instruct models with 1.5B, 3B, and 7B parameters~\cite{r104}. Each was quantized to 4 bits (Q4\_K\_M) and served by llama.cpp~\cite{r9} with temperature~0 and at most 64 output tokens. Passages and queries were embedded with BGE-M3~\cite{r42} and indexed with HNSW~\cite{r43}. The top-$k$ passages went into the prompt, which instructed the model to answer only from the numbered passages, to cite a passage number, and otherwise to reply \texttt{UNANSWERABLE}. We ran five configurations: the three model sizes with $k=3$, and the 3B model with $k=1$ and $k=5$. Retrieval was computed once and reused, so every configuration saw identical passages and scores.

\emph{Knowledge pack.} The corpus comprised 950 passages from 28 of the 35 articles of the SQuAD~2.0 development set~\cite{r105}; the remaining 7 articles were held out to create knowledge gaps. The corpus and its index were packaged with a manifest of SHA-256 hashes signed with Ed25519~\cite{r106}. At activation, the device verified the signature and every hash before loading the pack. Two tamper tests checked that a modified passage and an unsigned manifest edit are both rejected.

\emph{Queries.} We sampled 300 questions with a fixed seed, 100 per condition. \emph{Answerable} questions have an answer in a passage of the pack. \emph{Unanswerable} questions are SQuAD~2.0's adversarially written questions that are labelled unanswerable for a passage that is in the pack. \emph{Out-of-corpus} questions are answerable questions about the held-out articles. The first condition should be answered; the other two should lead to abstention. These labels are defined relative to the original passage or article, not the whole pack, and we did not re-annotate them against all 950 passages. An unanswerable-labelled question could therefore be answerable from another passage, and a held-out question could have an answer elsewhere. As partial evidence, across all configurations there were only five non-empty, non-abstaining responses to out-of-corpus questions, covering four distinct questions, and none contained a normalized gold answer; answers to unanswerable-labelled questions were not checked against other passages.

\emph{Trust mechanisms.} We compared layered routing policies. The first is the generator's own \emph{self-abstention} (M). The second adds the deterministic \emph{output validator} (V), which treats a response containing only citation markers and no answer text as an abstention. The third adds a \emph{retrieval-score gate} (G), which abstains when the top-1 cosine score falls below a threshold. Thresholds follow split conformal calibration~\cite{r55}: in each of 2,000 random splits, the threshold is the $\lfloor\alpha(n+1)\rfloor$-th smallest top-1 score among the answerable queries of the calibration half ($\alpha=0.1$), and it is evaluated on the other half. When this rank is below one, we use a threshold that rejects no scores (zero for entailment probabilities and $-1$ for cosine similarity). Each split-specific threshold is a valid split-conformal threshold for its held-out half, with a marginal guarantee, under exchangeability, that at least 90\% of answerable queries pass the gate. This guarantee concerns retention of answerable queries, not correctness or selective risk. We report held-out results across splits, and, for the tables, apply the mean of the split thresholds to all queries ($\hat\tau=0.50$, standard deviation 0.02 across splits). The mean threshold is an empirically calibrated gate and carries no formal guarantee itself. For the full layered policies (M+V+G, and M+V+G with the entailment verifier described below), we also re-fitted both thresholds in each of the 2,000 splits and evaluated on the held-out half; the held-out means differed from the full-sample values in Tables~\ref{tab8} and~\ref{tab9} by at most 1.0 percentage points (risk: 0.009) for M+V+G and 2.1 percentage points (risk: 0.029) with the verifier. The tables therefore report full-sample empirical evaluation with mean thresholds. Paired bootstrap intervals keep the fitted thresholds fixed and so exclude threshold-estimation uncertainty. Their resampling units are questions, not articles, so they do not capture variability from a new corpus or article split, another model, or another device.

\emph{Metrics.} An answer counts as correct if it contains a normalized gold answer (\emph{lenient}); we also report token-F1\,$\ge$\,0.5 (\emph{strict}). \emph{Risk} is the fraction of answered queries that were not answered correctly. Separation between conditions is summarized by the AUROC with 95\% bootstrap intervals. Differences between configurations and policies are reported with paired bootstrap 95\% intervals over the same questions (4,000 resamples). A pilot with 210 queries, three held-out articles, and the 3B model gave consistent results; we report the larger study.

\emph{Timing and energy.} End-to-end latency is the sum of query embedding and retrieval, measured once per question in the shared retrieval pass, and generation, measured as the wall-clock time of the request to the model server (prompt processing and decoding). Validator time is negligible; verifier time is reported separately and is not included. Board power was sampled every 0.5~s from the Raspberry Pi~5 power-management IC by summing voltage times current over all rails it reports, without idle subtraction. Energy per query is estimated as the mean power during generation multiplied by the end-to-end latency. These are estimates, not external measurements. Configurations ran sequentially (1.5B, 3B with $k=3$, $k=1$, $k=5$, then 7B), with the model server restarted for each and its default in-memory prompt cache enabled; each question was asked once per configuration, and cache effects were not isolated. No warm-up queries were run.

\emph{Reproducibility.} We used the official Qwen2.5-Instruct GGUF files (Q4\_K\_M), llama.cpp commit 098ce52, temperature~0, seed~42, a 4,096-token context, one server slot, and four threads; HNSW used $M=16$, construction depth 200, and search depth 100. SHA-256 digests of all model files and the Hugging Face revisions of BGE-M3 and the NLI checkpoint are recorded in the release. The question sample, prompts, and all scripts are released (Data and Code availability).

\emph{Post-hoc entailment verifier (RQ2).} For every non-empty answer, a natural language inference (NLI) cross-encoder, DeBERTa-v3-base fine-tuned on SNLI and MultiNLI~\cite{r107,r108}, scored whether each retrieved passage entails the answer. Following work on verifying question answering with NLI~\cite{r109,r110}, each question--answer pair was first converted into a declarative statement with rule-based question-to-declarative rewriting (for example, \emph{What is Warsaw's symbol?} with answer \emph{the mermaid} becomes \emph{Warsaw's symbol is the mermaid}). The verifier abstains when the maximum entailment probability falls below a threshold calibrated, as for the gate, to retain at least 90\% of correct answers. We also recorded a deterministic \emph{span check}: whether the normalized answer occurs as a normalized span of a retrieved passage. The verifier was applied post hoc to all 1,500 generations; generation was not repeated. We used the \texttt{cross-encoder/nli-deberta-v3-base} checkpoint, took the entailment label index from its configuration, and truncated passages (not the hypothesis) at 512 tokens.

\emph{Update-security tests (RQ4).} We extended the pack manifest with a pack identifier, a monotonic sequence number, an expiry time, the embedding model and dimension, and a signing-key identifier, in the spirit of TUF and the IETF IoT update manifest~\cite{r85,r86}. The device holds a trust store (current publisher key and a revoked earlier key) and its installed sequence number. We compared a \emph{signature-and-hash} baseline (a signature by any trusted key, plus file hashes, as in the main experiment) with a \emph{layered} verifier that verifies the signature only with the key declared in the manifest, which must be in the trust store and not revoked, and also enforces a strictly increasing sequence number, expiry against the device clock, and compatibility with the installed embedding model. Binding signature verification and revocation to the same key matters: our initial implementation verified the signature with any trusted key but checked revocation against the declared key identifier, so it accepted a manifest signed with the revoked key that declared the current key's identifier. We found this flaw during review, corrected it, added the case to the test set, and repeated all update-security tests; we report the corrected results. Both were tested on eleven packs: a valid newer pack, nine attack variants, and the poisoned pack described next.

\emph{Signed poisoning (RQ5).} Following PoisonedRAG~\cite{r78}, we selected 30 answerable questions that the 3B model had answered correctly and, for each, wrote five short passages that restate the question and assert an incorrect answer drawn from another question of the same type. The 150 passages were added to the corpus, embedded on the device, and released as a newer pack validly signed by the publisher key. We then repeated retrieval and generation (3B, $k=3$) for the 30 targets. An attack succeeds when the answer contains the injected answer and not the gold answer.

\section{Results}\label{sec:results}

\begin{table*}[!t]
\caption{Proof-of-concept results on a Raspberry Pi~5 under network isolation approximating C4, with 100 queries per condition. Correct: share of the 100 answerable questions answered correctly (lenient/strict). Empty: share of all 300 queries answered with citation markers but no answer text. M: self-abstention; M+V: with output validator; M+V+G: with validator and calibrated score gate (mean split-conformal threshold 0.50, $\alpha=0.1$). Latency and energy are end-to-end estimates (Sect.~\ref{sec:method}). Risk: fraction of answered queries not answered correctly}\label{tab8}
\tablebodyfont\setlength{\tabcolsep}{3pt}
\begin{tabular}{@{}lccccccccc@{}}
\toprule
Configuration & Correct & Empty & \multicolumn{2}{c}{Unanswerable abstained} & \multicolumn{3}{c}{Risk among answered} & Median & Energy \\
\cmidrule(lr){4-5}\cmidrule(lr){6-8}
 & len./strict & & M & M+V+G & M & M+V & M+V+G & latency & per query \\
\midrule
Qwen2.5-1.5B, $k=3$ & 0.27/0.17 & 0.30 & 0.47 & 0.89 & 0.81 & 0.46 & 0.43 & 10~s & 77~J \\
Qwen2.5-3B, $k=1$ & 0.45/0.19 & 0.00 & 0.80 & 0.84 & 0.44 & 0.44 & 0.42 & 8~s & 62~J \\
Qwen2.5-3B, $k=3$ & 0.75/0.31 & $<$0.01 & 0.66 & 0.72 & 0.39 & 0.38 & 0.37 & 22~s & 174~J \\
Qwen2.5-3B, $k=5$ & 0.52/0.20 & $<$0.01 & 0.68 & 0.75 & 0.49 & 0.49 & 0.44 & 36~s & 286~J \\
Qwen2.5-7B, $k=3$ & 0.75/0.74 & 0.04 & 0.71 & 0.84 & 0.36 & 0.29 & 0.24 & 50~s & 399~J \\
\midrule
\multicolumn{10}{@{}p{0.97\textwidth}@{}}{\emph{Retrieval (all configurations):} recall@1/3/5 = 0.83/0.98/0.98; gate AUROC vs.\ out-of-corpus 0.97 [0.94, 0.99], vs.\ unanswerable 0.68 [0.60, 0.76]}\\
\multicolumn{10}{@{}p{0.97\textwidth}@{}}{\emph{Out-of-corpus abstained:} 0.91--1.00 under all policies with self-abstention; 0.99--1.00 with M+V+G}\\
\multicolumn{10}{@{}p{0.97\textwidth}@{}}{\emph{Knowledge pack:} passage embedding 28.8~min on device; verify 4.4~ms (median of 20); load 2.7~s; tampered passage and unsigned manifest edit rejected}\\
\botrule
\end{tabular}
\end{table*}

\subsection{RQ1: Retrieval-Score Gating}\label{sec:res-gating}

\begin{figure*}[!t]
\centering
\includegraphics[width=\textwidth]{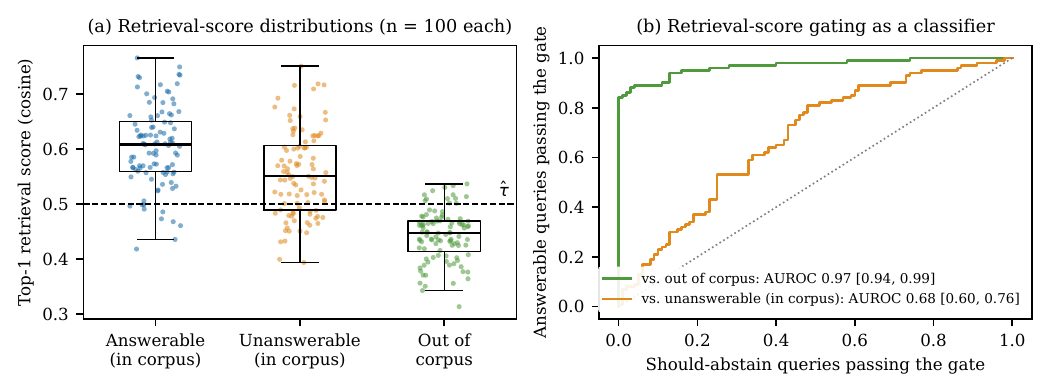}
\caption{Retrieval-score gating on the edge device (100 queries per type). (a) Distribution of top-1 cosine scores per query type; the dashed line marks the mean calibrated threshold $\hat\tau$. (b) ROC curves of the gate for separating answerable queries from out-of-corpus and from in-topic unanswerable queries, with AUROC and 95\% bootstrap intervals}\label{fig3}
\end{figure*}

The gold passage was retrieved at rank~1 for 83\% and within the top~3 for 98\% of answerable questions. Mean top-1 scores were 0.60 (SD 0.07) for answerable, 0.56 (0.08) for in-topic unanswerable, and 0.44 (0.05) for out-of-corpus queries (Fig.~\ref{fig3}a). The gate therefore separated answerable from out-of-corpus queries almost perfectly, with AUROC 0.97 [0.94, 0.99], but separated them poorly from in-topic unanswerable queries, with AUROC 0.68 [0.60, 0.76] (Fig.~\ref{fig3}b). On held-out halves, the split-specific thresholds retained on average 91\% of answerable queries (2.5th percentile 77\%; at least 90\% in 61\% of splits), consistent with the marginal guarantee, and rejected 88\% of out-of-corpus queries but only 28\% of in-topic unanswerable ones. Retrieval-side protection is thus conditional, as the framework anticipates (Sect.~\ref{sec:trust}): a query about a topic the corpus covers can be highly similar to relevant passages and still be unsupported. Because the out-of-corpus questions come from seven held-out articles, this separation reflects one article split and should be read as conditional on it.

\subsection{RQ2: Layered Trust Mechanisms}\label{sec:res-layers}

\emph{Self-abstention handles knowledge gaps but not in-topic unanswerable questions.} All models declined almost all out-of-corpus questions (91--99\%). They declined far fewer in-topic unanswerable ones: 47\%, 66\%, and 71\% for the 1.5B, 3B, and 7B models (Table~\ref{tab8}).

\begin{figure*}[!t]
\centering
\includegraphics[width=\textwidth]{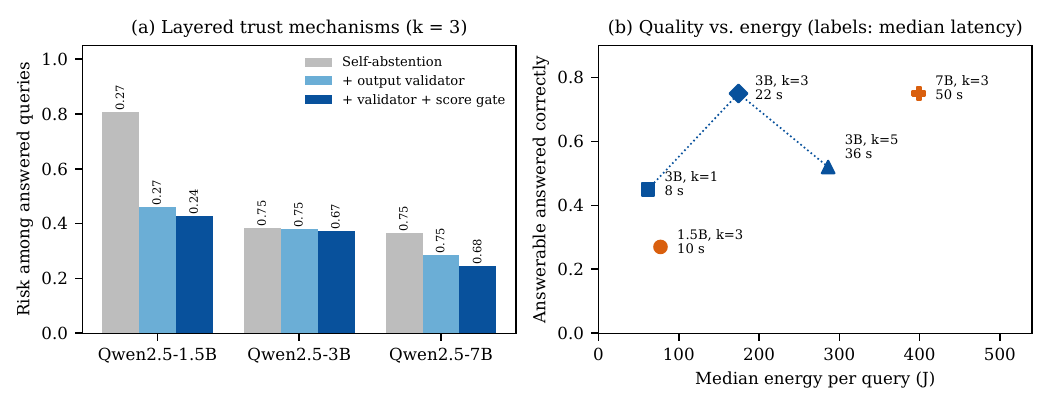}
\caption{Layered trust mechanisms and the quality--cost design space. (a) Risk among answered queries for three model sizes ($k=3$) under self-abstention, with an output validator, and with validator and score gate; the numbers above the bars give the fraction of answerable queries answered correctly. (b) Fraction of answerable queries answered correctly against median energy per query for all five configurations; labels give median latency}\label{fig4}
\end{figure*}

\emph{Small models fail by emitting empty outputs, which a deterministic validator turns into explicit abstentions.} The 1.5B model frequently responded with a citation marker and no answer text: this happened for 30\% of all queries, including 44\% of answerable and 39\% of unanswerable ones. The 3B model did so for 0.3\% of queries and the 7B model for 4\%. Treating such outputs as abstentions cost no correct answers in our sample. It lowered the risk among answered queries from 0.81 to 0.46 for 1.5B (paired 95\% interval of the change $-0.46$ to $-0.25$) and from 0.36 to 0.29 for 7B ($-0.13$ to $-0.04$), and raised the share of unanswerable questions declined to 86\% and 77\%, respectively (Fig.~\ref{fig4}a). This improvement re-routes malformed outputs rather than reducing substantive false claims. Over all 300 queries, the 1.5B model gave 27 correct and 23 incorrect substantive answers, 160 explicit abstentions, and 90 empty outputs; the 7B model gave 75 correct and 30 incorrect substantive answers, 182 abstentions, and 13 empty outputs.

\emph{The score gate adds a consistent but costly increment.} Adding the gate on top of self-abstention and the validator raised abstention on in-topic unanswerable questions by 3--7 percentage points (3B: $+6$, paired 95\% interval $+2$ to $+11$; 7B: $+7$, $+3$ to $+13$) and on out-of-corpus questions to 99--100\%. It reduced risk modestly, for example from 0.29 to 0.24 for 7B ($-0.09$ to $0.00$). In exchange, it removed 2--8 percentage points of correct answers (3B: $-8$, $-14$ to $-3$; 7B: $-7$, $-13$ to $-2$). The increment was similar across model sizes. We had expected larger models to be more overconfident and therefore to gain more from the gate; in this setting, the larger model was the more cautious one. The gate's distinct value lies in being independent of the generator and in having an explicit, empirically validated retention target for answerable queries. Even the strongest layered policy (7B with validator and gate) declined only 84\% of in-topic unanswerable questions.

\emph{The entailment verifier we tested adds little.} It separated correct from incorrect answers only weakly, with AUROC between 0.50 and 0.64 across configurations (Table~\ref{tab9}). The calibrated threshold collapsed to near zero, because about a tenth of correct answers received entailment probabilities below 0.001. Two implementation factors contribute. The rule-based rewriting fell back to a generic template (\emph{The answer to the question "\ldots" is \ldots}) for 26--35\% of answers and produced some ungrammatical statements, and the small NLI model is brittle on such inputs. Of the correct answers with entailment below 0.01 (22 for 3B and 24 for 7B), 6 and 7 used the fallback template; we did not further separate rewriting errors from classification errors. Added on top of the other layers, the verifier raised abstention on in-topic unanswerable questions by 0--5 percentage points (3B: $+5$, $+1$ to $+10$; 7B: $+3$, $0$ to $+7$) at a cost of 3--7 percentage points of correct answers (3B: $-5$, $-10$ to $-1$; 7B: $-7$, $-12$ to $-3$) and a median 2.1--2.4~s per answer at $k=3$. The span check shows why lexical support checks cannot help here. For the 7B model, 96\% of its substantive answers to in-topic unanswerable questions (22 of 23) matched a normalized span of a retrieved passage, the same rate as for its correct answers; for the 3B model ($k=3$) the share was 47\%. These outputs match passage spans despite being scored as incorrect under the benchmark labels, so a span check cannot separate them; a manual review of such answers would strengthen this interpretation. Whether a more faithful question-conditioned entailment check could is not settled by our results: prior work shows that NLI can flag answers that do not address the question~\cite{r109}, and our verifier's weakness is confounded by its rewriting and model size.

\begin{table*}[!t]
\caption{Post-hoc entailment verification (E) added to self-abstention, output validator, and score gate (M+V+G). AUROC: separation of correct answers (containing a gold answer) from incorrect ones by maximum entailment probability; ``incorrect'' refers to gold-answer correctness, not to evidence non-entailment. Span present: share of answers matching a normalized span of a retrieved passage, for correct answers and for answers given to in-topic unanswerable questions}\label{tab9}
\tablebodyfont\setlength{\tabcolsep}{2.4pt}
\begin{tabular}{@{}lcccccc@{}}
\toprule
Configuration & AUROC [95\% CI] & Unanswerable abstained & Correct & Risk & Span present & Verify time \\
 & & M+V+G $\rightarrow$ +E & M+V+G $\rightarrow$ +E & M+V+G $\rightarrow$ +E & correct / unans. & (median) \\
\midrule
Qwen2.5-1.5B, $k=3$ & 0.50 [0.34, 0.66] & 0.89 $\rightarrow$ 0.89 & 0.24 $\rightarrow$ 0.20 & 0.43 $\rightarrow$ 0.44 & 0.56 / 0.64 & 2.2~s \\
Qwen2.5-3B, $k=1$ & 0.60 [0.47, 0.72] & 0.84 $\rightarrow$ 0.87 & 0.42 $\rightarrow$ 0.39 & 0.42 $\rightarrow$ 0.40 & 0.36 / 0.55 & 0.8~s \\
Qwen2.5-3B, $k=3$ & 0.64 [0.54, 0.73] & 0.72 $\rightarrow$ 0.77 & 0.67 $\rightarrow$ 0.62 & 0.37 $\rightarrow$ 0.35 & 0.43 / 0.47 & 2.4~s \\
Qwen2.5-3B, $k=5$ & 0.55 [0.44, 0.66] & 0.75 $\rightarrow$ 0.77 & 0.50 $\rightarrow$ 0.47 & 0.44 $\rightarrow$ 0.41 & 0.42 / 0.59 & 4.4~s \\
Qwen2.5-7B, $k=3$ & 0.55 [0.43, 0.66] & 0.84 $\rightarrow$ 0.87 & 0.68 $\rightarrow$ 0.61 & 0.24 $\rightarrow$ 0.24 & 0.96 / 0.96 & 2.1~s \\
\botrule
\end{tabular}
\end{table*}

\subsection{RQ3: Model Size and Context Size}\label{sec:res-size}

\emph{Model size: 3B offers a lower-cost trade-off.} With $k=3$, the share of answerable questions answered correctly under the lenient criterion was 27\% for 1.5B, 75\% for 3B, and 75\% for 7B. The 3B and 7B models disagreed on 32 answerable questions, 16 in each direction; the paired 95\% interval for the difference is $-12$ to $+11$ percentage points, so equal point estimates on 100 questions do not establish equivalence. On the strict criterion, the 7B model was far better (74\% versus 31\%; difference $+43$, $+30$ to $+56$), largely because it answered concisely (a median of 2 answer tokens against 12), although containing the gold span does not exclude incorrect or irrelevant material in a verbose answer. Under the validator and gate, the 7B model also had lower risk (0.24 versus 0.37; difference $-0.13$, $-0.22$ to $-0.04$) and declined more in-topic unanswerable questions. For this, it needed 2.3 times the latency (median 50~s versus 22~s) and estimated energy (399~J versus 174~J), and its server's resident memory after the first completed query was 8.7~GiB versus 3.8~GiB.

\emph{Context size: more retrieved passages hurt.} For the 3B model, recall was 83\% at $k=1$ and 98\% at both $k=3$ and $k=5$. Yet answer quality peaked at $k=3$ (75\% correct) and fell at $k=5$ (52\%; paired difference $-23$ percentage points, $-32$ to $-14$), while median latency rose from 22~s to 36~s and energy from 174~J to 286~J (Fig.~\ref{fig4}b). With five passages, the model declined more answerable questions (31\% versus 13\% at $k=3$) and answered less precisely (among answered, answerable-labelled queries, 75\% of its answers were correct, versus 86\% at $k=3$). At $k=1$ it was fastest (8~s, 62~J) but too conservative: it declined 39\% of answerable questions, and only 45\% were answered correctly. Two effects are at work. The rising latency and energy reflect computational cost, since prompt length grows with $k$. The falling precision at $k=5$, with recall unchanged, is plausibly explained by distraction from the additional passages, although our design does not isolate this cause.

\emph{Cost profile.} Prompt processing dominated generation time: it accounted for 89--95\% of model time at $k\ge3$ and 71\% at $k=1$. Prompt throughput fell from 58 to 28 to 11 tokens/s, and decoding from 11.1 to 5.6 to 2.6 tokens/s, for the 1.5B, 3B, and 7B models. Query embedding and retrieval took a median of 0.72~s. Mean board power stayed between 7.6 and 8.0~W in all configurations, so energy per query scaled with latency. The model server's resident memory grew during each run. The maximum resident set size sampled at query completion was 6.3, 9.4, and 13.6~GiB for the three model sizes at $k=3$, and 5.8 and 11.8~GiB for the 3B model at $k=1$ and $k=5$; the contribution of the configured prompt cache was not isolated experimentally.

\subsection{RQ4: Knowledge-Pack Activation}\label{sec:res-packs}

Embedding the 950 passages alone took 28.8~min on the device; index construction and packaging were not separately timed. In contrast, signature and hash verification took a median of 4.4~ms over 20 repetitions (maximum 6.4~ms), including reading and hashing both pack files from a warm page cache, and loading the pack together with the embedding model took about 2.7~s. Both the modified passage (hash mismatch) and the unsigned manifest edit (invalid signature) were rejected. The asymmetry, roughly five orders of magnitude between verification and passage embedding, and about three between loading and passage embedding, supports the operational model of Sect.~\ref{sec:packs}: in C3--C4 regimes, publishers build and sign packs off-device, and edge nodes verify and activate them. \emph{Signature and hash checks alone stop only integrity attacks.} Table~\ref{tab10} shows the outcome of the update-security tests. The signature-and-hash baseline rejected the three integrity attacks: a manifest that no trusted key verifies (unknown signer), a modified passage, and an index from a different build. It accepted six of the remaining attacks: an older pack (rollback), a re-issued current version (replay), an expired pack, a pack built for a different embedding model, a pack signed with a revoked key, and a pack signed with the revoked key whose manifest declares the current key's identifier. The layered verifier rejected all nine attacks and accepted the valid newer pack, with similar measured latency (median 4.55~ms versus 4.53~ms for the baseline on the valid pack). Expiry checking relies on the device clock. Without a trusted time source, sequence numbers still prevent acceptance of packs older than the preserved state, but cannot detect that newer packs are being withheld; they complement rather than replace freshness information~\cite{r86}. These tests demonstrate rejection of the specified invalid packs, not comprehensive update security. They assume that the installed sequence state cannot be reset by an attacker, that trust-store and revocation updates are themselves authenticated, that version state is scoped to the pack identifier, that the verified bytes are the bytes activated, and that activation and state persistence are crash-safe; none of these was tested.

\subsection{RQ5: Validly Signed Poisoning}\label{sec:res-poison}

Both verifiers accepted the poisoned pack, as expected: it was newer, unexpired, compatible, and signed by the current publisher key. The 30 targets were, by construction, questions the 3B model had answered correctly on the clean pack, and the attack added 150 passages, about 15.8\% of the original 950. The poisoned passages were retrieved for all 30 targets and occupied on average 2.9 of the 3 context slots. Correct answers on the targets fell from 30 of 30 to 0 of 30. The model gave the injected answer for 14 targets (47\%), abstained for 15, and gave another answer for one; the outcome thus combines answer corruption with loss of availability. Score gating offered no protection: poisoned passages restate the question, so their mean top-1 score (0.80) far exceeded the threshold (0.50). Post-hoc verification did not help either: the injected answers received a mean entailment probability of 0.92 from the passages that asserted them, and adding the span and entailment checks lowered attack success only from 14 to 13 targets; the one rejection came from the span check, and the entailment threshold rejected none of the 14. This demonstrates, on an edge system and under the tested budget, the known content-trust problem of RAG~\cite{r78}; it does not establish a minimum poisoning budget. The mechanisms we evaluated check different properties (output format, retrieval similarity, lexical or entailment support, and artifact authenticity and integrity), but none of them establishes whether the retrieved evidence is true. For signed packs, content trust must therefore be established before signing, through curation, provenance, and adversarial testing.

\begin{table*}[!t]
\caption{Update-security tests and validly signed poisoning on the Raspberry Pi~5. Signature + hash: signature by any trusted key plus file hashes. Layered: signature verified only with the declared, non-revoked key, plus file hashes, strictly increasing sequence number, expiry, and embedding-model compatibility. Median verification time for the valid pack: 4.53~ms (signature + hash) and 4.55~ms (layered); results from the corrected layered verifier}\label{tab10}
\tablebodyfont\setlength{\tabcolsep}{4pt}
\begin{tabular}{@{}lllp{5.2cm}@{}}
\toprule
Pack presented to the device & Signature + hash & Layered & Reason for rejection (layered) \\
\midrule
Valid newer pack & accept & accept & -- \\
Older pack (rollback) & \textbf{accept} & reject & sequence not newer \\
Current version re-issued (replay) & \textbf{accept} & reject & sequence not newer \\
Expired pack & \textbf{accept} & reject & expired \\
Built for a different embedding model & \textbf{accept} & reject & incompatible embedding model \\
Signed with a revoked key & \textbf{accept} & reject & declared key revoked \\
Revoked key, manifest declares current key & \textbf{accept} & reject & signature fails with declared key \\
Signed with an unknown key & reject & reject & declared key not in trust store \\
Modified passage & reject & reject & hash mismatch (corpus) \\
Index from a different build & reject & reject & hash mismatch (index) \\
Validly signed, poisoned content & \textbf{accept} & \textbf{accept} & -- \\
\midrule
\multicolumn{4}{@{}p{0.97\textwidth}@{}}{\emph{Poisoned pack, 30 targets (Qwen2.5-3B, $k=3$):} poison in top-3 for 30/30 targets (mean 2.9 of 3 slots); correct answers 30/30 before, 0/30 after; attack success 47\% with no defense, 47\% after score gating (mean top-1 score 0.80, $\hat\tau=0.50$), 43\% after span and entailment checks (mean entailment of injected answers 0.92)}\\
\botrule
\end{tabular}
\end{table*}

\section{Discussion}\label{sec:discussion}

\subsection{Design Guidelines}\label{sec:guidelines}

Six guidelines follow for generative edge deployments under the conditions studied.
\begin{enumerate}
\item \emph{Place the deployment on the C0--C4 axis first.} The regime determines which update and assurance mechanisms are admissible. At C3--C4, knowledge packs replace iterative federation as the update channel, and verification of signature, version state, and expiry at activation becomes the integrity anchor. Content trust must be established separately, through curation and testing.
\item \emph{Evaluate mid-sized models first.} On our Pi-class CPU, a 3B model achieved the same gold-answer containment rate as a 7B model at less than half the latency and estimated energy, while the 7B model answered more concisely and with lower risk. This is a configuration-specific trade-off, not a general law: the optimum depends on model family, runtime, task, and quality metric, so model size should be tuned against a quality, latency, and energy budget.
\item \emph{Tune the number of retrieved passages; do not maximize it.} Prompt processing dominates latency, and in our setting five passages lowered both correctness and precision relative to three.
\item \emph{Add deterministic output checks.} A trivial validator converted malformed, citation-only outputs into explicit abstentions with negligible overhead and no observed loss of correct answers, which matters most for small models.
\item \emph{Use score gating for knowledge gaps, and do not expect support checks to catch in-topic unanswerable questions.} The gate detects out-of-corpus queries almost perfectly, and split-conformal calibration gives it a validated retention target for answerable queries. In-topic unanswerable questions, however, are often answered with spans that match the retrieved evidence, and self-abstention leaves a substantial share of them answered (29--34\% for the 3B and 7B models), and the gate and the span and entailment checks we tested improved on this only modestly. Answerability-aware verification or human review is needed for high-stakes use.
\item \emph{Enforce version state, not just signatures, and establish content trust before signing.} Layered verification of sequence numbers, expiry, key revocation, and compatibility had similar measured latency and stopped the tested update attacks that signature and hash checks missed. The signature checks and inference-time checks we evaluated did not detect the poisoned content in our attack, so curation, provenance, and adversarial testing of packs must happen at the publisher.
\end{enumerate}

\subsection{Implications for Governance Evidence}\label{sec:governance}

Every signal used in our pipeline can be recorded per query: pack version and verification outcome, retrieval scores, validator decision, gate decision, verifier score, and the cited passage. Our prototype logged pack verification outcomes, retrieval scores, answers, and verifier scores; validator and gate decisions are derived from these released logs. Such records reconstruct which knowledge was available at a given time, what evidence was retrieved, and why the system answered or abstained. These are the kinds of evidence that AI Act record-keeping and human-oversight obligations, where they apply, and the measurement and management functions of the NIST AI RMF call for~\cite{r5,r6,r94,r95}. Local deployment does not automatically satisfy GDPR data minimization, which is a property of the processing rather than of its location~\cite{r4}. But it makes evidence that is hard to obtain from cloud deployments tractable to collect.

\subsection{Threats to Validity}\label{sec:validity}

\emph{Internal validity.} Decoding was deterministic and retrieval was held fixed across configurations; sequential execution and cache effects remain potential confounders, particularly for cost comparisons. The lenient criterion can credit verbose answers that contain the gold span, and the strict criterion penalizes them; we report both. Power was estimated from on-board sensors rather than measured externally, without idle subtraction, and energy combines generation-time power with end-to-end latency. Configurations ran sequentially with a server-side prompt cache and no warm-up. The entailment verifier used a small CPU-friendly NLI model with rule-based question-to-declarative rewriting; larger NLI models, neural rewriting, or an LLM judge may verify better. The poisoning experiment used 30 targets and templated passages, which a curator might detect more easily than optimized attacks~\cite{r78,r79}.

\emph{External validity.} The study uses one device, one model family, one English Wikipedia-based dataset, and 100 queries per condition. SQuAD~2.0's unanswerable questions are adversarially constructed, and the out-of-corpus questions come from seven held-out articles, so absolute rates should not be generalized; one hundred questions from seven articles are clustered and do not amount to evidence from one hundred independent topics. Labels were not re-validated against the whole pack (Sect.~\ref{sec:method}). Domain corpora such as maintenance manuals or clinical guidelines, and other model families, may behave differently.

\emph{Construct validity.} For simplicity, the signing key was generated on the device, whereas a real C3--C4 deployment would keep it with an off-device publisher. Network isolation was enforced by a namespace rather than physical separation; the device's behavior is the same, but the threat model of a physical air gap is not exercised.

\subsection{Open Problems}\label{sec:open}

Our results sharpen several open problems.
\begin{itemize}
\item \emph{Integrated benchmarks.} Existing benchmarks measure either speed~\cite{r39} or RAG quality~\cite{r44,r45,r46}, but no widely adopted benchmark jointly measures quality, abstention, latency, energy, and trust under disconnected conditions.
\item \emph{Calibration under shift and richer unanswerability.} Conformal and certified methods for RAG~\cite{r26,r27,r28,r30,r56} were not designed for edge operation, and our results show that retrieval-side calibration alone cannot detect in-topic unanswerable questions. Evaluations should cover the full range of unanswerability categories~\cite{r29}, compare score margins, rerankers, and entailment checks, and test calibration transfer across corpora.
\item \emph{Answerability-aware verification.} Lexical support checks cannot separate correct answers from incorrect ones that match spans of the evidence, and the entailment verifier we tested separated them only weakly. Edge-feasible verifiers that test whether the evidence answers the question asked, with faithful question-to-statement conversion or stronger models, remain an open problem.
\item \emph{Content trust for signed knowledge.} Signed packs can carry poisoned content that passes every cryptographic and inference-time check. Publisher-side curation, provenance attestation, and adversarial pack testing need to become part of the update process.
\item \emph{Multilingual evaluation.} Multilingual RAG benchmarks are emerging~\cite{r48}, and multilingual embeddings make non-English local RAG feasible~\cite{r42}, but edge evaluations for less-resourced languages remain scarce.
\item \emph{Freshness under disconnection.} Temporal RAG models evolving knowledge and incremental corpus updates~\cite{r111}; evaluating such methods under controlled offline update windows, including staleness bounds and their communication to users, remains open.
\item \emph{Cross-air-gap update protocols.} Model signing~\cite{r83} and secure update frameworks~\cite{r85,r86} provide the building blocks. We found, however, no RAG-specific profile that jointly covers corpus and index provenance, sequence state, expiry, revocation, and verification at activation, and no peer-reviewed study of rollback or freshness attacks on RAG corpora. Our layered verifier shows that the necessary checks are cheap on the device, but a standardized profile is missing.
\item \emph{Offline knowledge federation and intermittent adaptation.} Federated RAG~\cite{r68,r69,r70,r71} and federated instruction tuning~\cite{r63,r64} presuppose connectivity; offline, multi-producer pack federation and opportunistic adapter exchange at C2--C3 are largely unexplored.
\item \emph{Human factors.} Users may over-rely on AI recommendations even when explanations are provided~\cite{r112}; how citations, caveats, and abstentions shape reliance offline is unmeasured.
\end{itemize}

\section{Conclusion}\label{sec:conclusion}

Trustworthy generative intelligence does not have to be cloud intelligence. We proposed a connectivity-regime framework that turns disconnection into a first-class design dimension, a taxonomy of inference-time trust mechanisms, and signed knowledge packs as the update channel for disconnected regimes. A network-isolated evaluation on a Raspberry Pi~5 then tested these ideas. Retrieval-score gating detected knowledge gaps almost perfectly but in-topic unanswerable questions poorly. A deterministic output validator turned malformed outputs into explicit abstentions with negligible overhead. A 3B model achieved the same gold-answer containment rate as a 7B model at less than half the cost, although the 7B model answered more concisely and with lower risk, and more retrieved context lowered answer quality. Verifying a signed pack took milliseconds, whereas embedding its passages on the device took about half an hour, and layered verification of version state stopped the tested update attacks that signature and hash checks missed. Yet a validly signed poisoned pack passed every pack check and eliminated correct answers on the targeted questions, and the entailment verifier we tested caught neither poisoned answers nor most answers to in-topic unanswerable questions. None of the mechanisms we evaluated establishes the truth of the evidence, and the support checks we tested did not reliably capture answerability; the remaining trust gaps lie there. For designers of intelligent environments that must keep working when the cloud is unreachable, these results offer concrete guidance on what to deploy, how to update it, and where the remaining trust gaps lie.

\bmhead{Abbreviations}
AI, artificial intelligence; AUROC, area under the receiver operating characteristic curve; AWQ, activation-aware weight quantization; CRA, Cyber Resilience Act; DP, differential privacy; FL, federated learning; GDPR, General Data Protection Regulation; IoT, Internet of Things; KV, key--value (cache); LLM, large language model; LoRA, low-rank adaptation; NIST AI RMF, NIST Artificial Intelligence Risk Management Framework; NPU, neural processing unit; OMS, OpenSSF Model Signing; OT, operational technology; PEFT, parameter-efficient fine-tuning; PKI, public-key infrastructure; QA, question answering; RAG, retrieval-augmented generation; TEE, trusted execution environment.

\section*{Statements and Declarations}

\bmhead{Use of AI tools}
The authors used standard writing and production tools (Grammarly, Overleaf) for language editing and \LaTeX{} preparation. Generative AI assistants were used to help identify candidate literature, to cross-check bibliographic metadata, and to help write the experiment and analysis scripts; every reference was verified by the authors against the publisher, DOI, or preprint-server record, and all code and results were reviewed by the authors. The AI tools did not generate the framework, the findings, or the conclusions. The authors take full responsibility for the manuscript.

\bmhead{Funding}
No funding was received to assist with the preparation of this manuscript.

\bmhead{Competing interests}
The authors are affiliated with securagen.ai P.C., which develops commercial software in the areas discussed in this paper. The framework, analysis, and conclusions were prepared independently of any specific product, and the experiment used open-source software and publicly available model weights under their respective licences. The authors have no other relevant financial or non-financial interests to disclose.

\bmhead{Ethics approval and consent to participate}
Not applicable.

\bmhead{Consent for publication}
Not applicable.

\bmhead{Data availability}
The per-query results, system information, and analysis outputs of the experiment (Sect.~\ref{sec:method}) are available at \repourl. The SQuAD~2.0 dataset is publicly available~\cite{r105}.

\bmhead{Materials availability}
Not applicable.

\bmhead{Code availability}
The code used to build the knowledge pack, run the network-isolated benchmark, and analyze the results is available at \repourl. The implementation uses open-source software (llama.cpp commit 098ce52, sentence-transformers 6.1.0, hnswlib 0.8.0, and PyTorch 2.14.1). Model weights are used under their respective licences, including the Qwen Research licence for Qwen2.5-3B-Instruct; no proprietary software of the authors' company was used.

\bmhead{Author contribution}
Georgios Feretzakis and Alexandros Papaspyridis had the idea for the article. Georgios Feretzakis designed the framework and the methodology, designed and conducted the experiment, and supervised and administered the project. Alexandros Papaspyridis critically revised the work. Both authors read and approved the final manuscript.

{\raggedright\setlength{\parskip}{0pt}

}

\end{document}